# A Technology for BigData Analysis Task Description using Domain-Specific Languages

Sergey V. Kovalchuk[1], Artem V. Zakharchuk[1], Jiaqi Liao[1],
Sergey V. Ivanov[1], Alexander V. Boukhanovsky[1,2]

[1]*ITMO University, Saint-Petersburg, Russia*
[2]*Netherlands Institute for Advanced Study in the Humanities and Social Sciences, Wassenaar, Netherlands*

kovalchuk@mail.ifmo.ru, zakharchukart@gmail.com, liaojiaqi08@gmail.com,
svivanov@mail.ifmo.ru, boukhanovsky@mail.ifmo.ru

**Abstract**
The article presents a technology for dynamic knowledge-based building of Domain-Specific Languages (DSL) to describe data-intensive scientific discovery tasks using BigData technology. The proposed technology supports high level abstract definition of analytic and simulation parts of the task as well as integration into the composite scientific solutions. Automatic translation of the abstract task definition enables seamless integration of various data sources within single solution.

*Keywords:* domain-specific language, big data, cloud computing,

## 1 Introduction

Today data-intensive scientific discovery is one of the most important issues within e-Science area (Hey, Tansley, & Tolle, 2009). There are a lot of data sources which can be considered as a subject to scientific discovery: data produced by simulation, obtained from various sensors, collected during observation, crawling etc. Nowadays such data forms large volume which can be mined to obtain information and knowledge. Combined with contemporarily available computational powers and volume of storage this leads to the popularity of such area as BigData (Philip Chen & Zhang, 2014) which forms specific approaches to management of data characterized by large volume, variety and velocity (as one of the definitions of BigData). Within this area a set of data processing models and technologies are developing to support growing requirements of current science and business tasks. One of the most popular data processing models is MapReduce (Doulkeridis & Nørvåg, 2013) used to organize distributed data processing without moving large amount of data over the network. There are a lot of solutions implementing this model for BigData tasks (one of the most popular among them is Hadoop (Apache Hadoop)). Within scientific tasks it is often required to combine the processing of large amount of data with complex simulation. This leads to integration of BigData solutions into existing WF-based platforms (see e.g. (Baranowski, Belloum, & Bubak, 2013)) and extension of existing concept of WF with BigData features. Nevertheless contemporary WF solutions are trying to make a shift towards a) more user-friendly approach which allows hiding WF structure behind the abstract structures more specific to the end user (which is often a domain specialist with a lack of programming skills) (McPhillips, Bowers, Zinn, & Ludäscher, 2009); b) system-level exploration instead of procedure-style calling services within WF (Foster & Kesselman, 2006). To support this shift there is a requirement of the new high-level expressive tools for data analytics task description in simple (for the user) terms (Assuncao, Calheiros, Bianchi, Netto, & Buyya, 2013). Such expressive tools can be integrated seamlessly within high-level simulation solutions to provide the user with the power of BigData hiding the complexity of that technology. On the other hand, one of the important issues in BigData area is integration and aggregation of data from various sources. Resolution of this issue requires implementation of high-level management of complex data structures (see e.g. (Fiore, D'Anca, Palazzo, Foster, Williams, & Aloisio, 2013), (Apache OODT)). It is much more important in case the automation of data analytics is implemented, since the data processing should be semantically integrated with the simulation process. In the presented work we are trying to implement knowledge-based technology in order a) to provide the user with ability to describe data analytics tasks using simple domain-specific terms; b) to support automatically the aggregation of data from various sources; c) to enable the integration of simulation-based e-Science solutions with BigData analytics tasks.

## 2  BigData within e-Science Tasks

Considering e-Science tasks and contemporary technologies used to build corresponding solutions several requirements can be proposed for BigData technology within this area:

1. *Integration of Various data sources.* Today there are a great diversity of data sources containing datasets different by the structure, format, origin (forecasts, estimations, measurements etc.), access protocol and veracity (the last one is often included into the definition of BigData). All these data should be accessed according to its nature and semantic meaning within the e-Science solutions. In the same time formatting, accessing and structural decomposition's specifics should be hidden as well as technological aspects of distributed data processing.

2. *Integration with simulation process.* The data analytics' tasks should be integrated with simulation tasks in two ways. First, they can be considered as a part of composite scientific applications used for simulation. One of the ways to perform this is extension of WF structure with specific nodes calling data analytics subroutines. Second, the task may require local simulation tasks to be solved during the data analysis (e.g. for classification of the data of estimate additional characteristics). As additional complication of the task it can require local calls of software packages to perform some complex data processing (e.g. forecasting simulation).

3. *High-level user interaction.* To support the user during the task definition the developed technology should use domain-specific semantics to describe high-level task. This semantic can be used to build expressive languages with textual or graphical notation. Such languages allow building composition interfaces (more powerful with graphical notation) as well as parameter definition interfaces or interfaces for result representation (can be automatically designed using domain-specific description).

4. ***Complex visualization.*** Large data visualization should support interactive exploration of data arrays with cognitive support and appropriate spatiotemporal scene rendering. Moreover the visualization should be tightly interconnected with simulation and data analysis tasks. To support this kind of data visualization in an automatic way the semantic description related to the data and interconnected processes should be used during the building of visual scene.

To support automatic task processing and data analytics integration a formalized domain specific knowledge can be used. This knowledge should include the description of the following artifacts:

1. ***Domain-specific semantics.*** A set of domain-specific objects considered within e-Science task should be described to define both the simulation process and the data analytics' procedures.

2. ***Data formats.*** The BigData technology should have the capability to work with high-level domain-specific abstract data types regardless of data storage specific. This requires either description of data formats or extraction procedures to be described.

3. ***Data aggregation patterns.*** To support distributed data processing with high-level abstraction the patterns for data structures construction and aggregation should be defined. This becomes more important in case the data chunks are distributed within the high-level data structures.

4. ***Integration with simulation infrastructure.*** The stored knowledge should include links to the simulation infrastructure to allow us a) to integrate the data analysis tasks into the composite applications; b) to use knowledge stored within the infrastructure to manage local calls of software packages.

Within the developed technology we propose a dynamic domain-specific language (DSL) (van Deursen, Klint, & Visser, 2000) to be considered as a knowledge-based expressive technology (Kovalchuk, Smirnov, Knyazkov, Zagarskikh, & Boukhanovsky, 2013) enabling implementation of semantically driven e-Science solutions which include WF-based simulation as well as distributed data processing.

# 3  DSL-based BigData Technology Implementation

## 3.1  Technological background

To implement the proposed technology a cloud computing platform ***CLAVIRE*** (Knyazkov, Kovalchuk, Tchurov, Maryin, & Boukhanovsky, 2012) was extended with data processing solution based on distributed data storage of the platform (Kovalchuk, Razumovskiy, & Spivak, 2013). The basic architecture of the distributed data processing solution is presented in the Figure 1.

The data processing solution implements fixed MapReduce scheme allowing us to process data files stored on the nodes of the distributed storage. The data processing module is implemented using Java programming language and uploaded to the storage with automatic replication onto the whole set of storage nodes. To run data processing task a request is processed by the core of data storage which accept configuration and additional parameters (query strings, additional data etc.) and run distributed processing of files with aggregation of the results. The distributed data processing solution is used as a basic technology which is extended by the mean of DSL integration. Within the storage architecture the DSL script is processed as a parameter of distributed task while the DSL interpreter is implemented as a data processing module and uploaded to the storage in a general way.

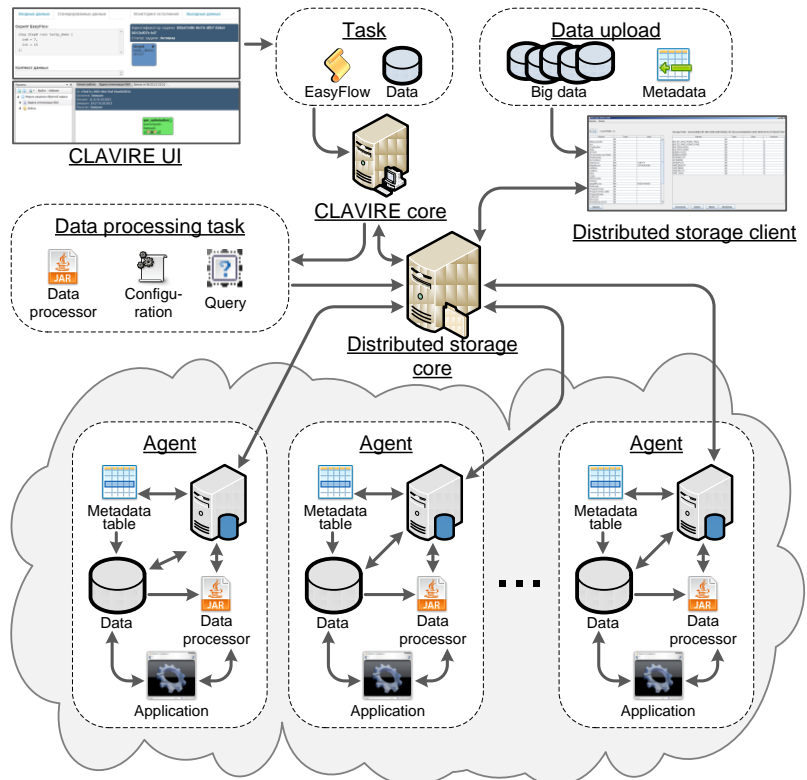

**Figure 1:** Distributed data processing within CLAVIRE platform

To support semantic integration and high-level concepts semantics the ***Virtual Simulation Objects (VSO)*** technology (Kovalchuk, Smirnov, Kosukhin, & Boukhanovsky, 2012) is used as one of the extensions of the CLAVIRE platform. The VSO concept and technology provide the high-level semantic abstractions which integrates domain-specific objective description of investigated system and simulation infrastructure. The integration with CLAVIRE enables the usage of abstract WFs in a form of EasyFlow (DSL for WF description) and high-level domain-specific parameters of available software packages described using EasyPackage (DSL for abstract software services description).

## 3.2 Implementation Details

The DSL developed within the proposed solution is based on Groovy language (Dearle, 2010), well integrated with Java language and virtual machine. The structure of developed DSL can be divided into several sets which define main groups of keywords and structures.

1. ***Domain objects.*** A set of objects and corresponding data structures is related to domain-specific semantics. This set is constructed using VSO technology and domain library.

2. ***Domain procedures.*** A set of procedures which implements domain-specific algorithms of data (domain objects) processing, extension and analysis. This set is constructed using domain library.

3. ***Domain-specific extensions.*** Define a set of keywords and shortcuts which makes the DSL shorter and easier to read. These structures are interpreted as procedures and methods during the DSL analysis. This set is constructed using domain library.

4. *Software description.* Is used for transform abstract local calls within data processing scripts into particular data files, parameters and calling commands. This set is constructed using EasyPackage description within PackageBase of CLAVIRE platform.

Each of these sets is related to several knowledge sources: VSO library, PackageBase (existing CLAVIRE extensions) and domain library (part of developing technology). Domain libraries by implementing fixed Java interface represent imperative (procedural) knowledge on acquisition of high-level structures from stored data and perform domain-specific analysis procedure on the data.

The request processed within the distributed storage (see Section 3.1) is extended with DSL query (a string containing all the statements is required for data processing and written using the developed DSL). The whole processing of the DSL query includes several steps (see Figure 2):

1. *DSL parsing.* Using the mentioned sources of knowledge the query is interpreted and translated into a set of task-specific procedures by the DSL interpreter with the use of extensions for particular domains. The procedure can include calls of domain-specific libraries, and local software calls.

2. *MapReduce operation.* Constructed procedures are used within map and reduce procedures which are performed within the distributed data storage. Map procedure processes files, while reduce procedure aggregates the results of processing. Additionally as the processing uses high-level data structures (which can be distributed across the nodes of storage) the processing of partially obtained data can be fulfilled only on the aggregation stage.

3. *Result presentation.* Aggregated results are presented with the links to the particular objects and their parameters (according to the VSO structure) and output parameters of software packages (according to the records in PackageBase). The results can be processed within the external composite application or presented to the end user.

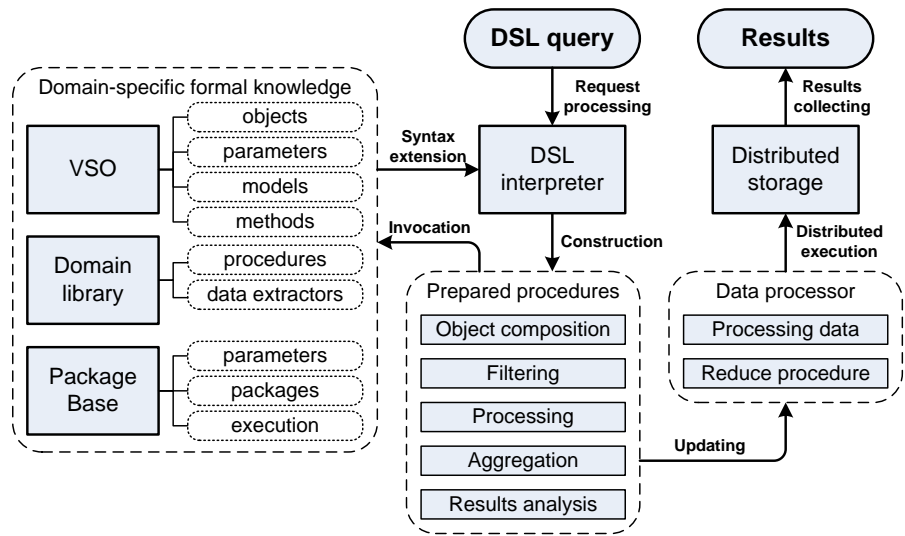

**Figure 2:** DSL query processing

The presented approach allows us to implement the hybrid software calling within composite application. According to the requirement 2 (see Section 2) the hybrid calling architecture (see Figure 3a) makes it possible both to call local software available on the agents of storage and to integrate distributed data processing into the WF structure as special kind of WF nodes. So the technology enables the usage of both data-to-code (regular WFs) and code-to-data (BigData) approaches within single composite application.

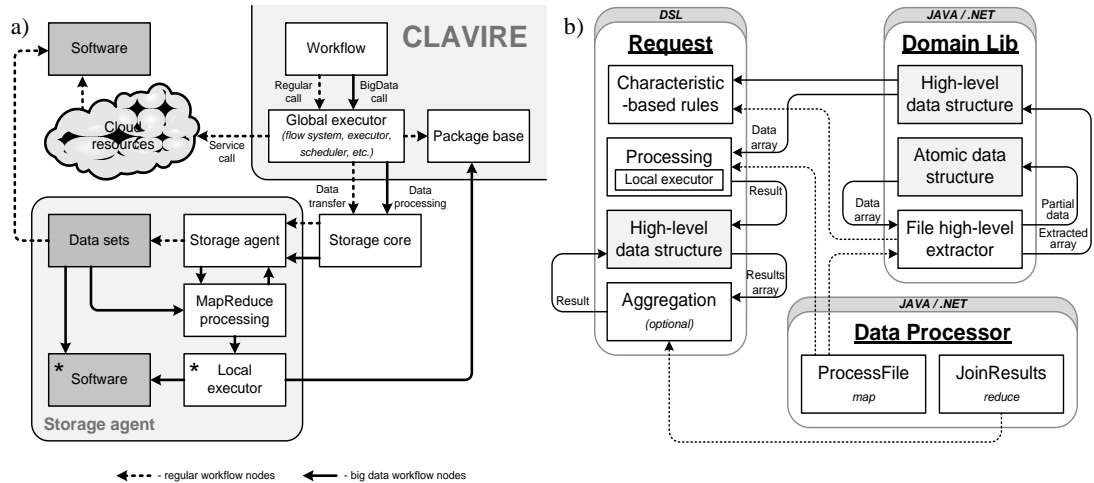

**Figure 3:** DSL-based data processing: a) hybrid software calling architecture;
b) high-level data structures processing

One of the important issues implemented within the developed technology is the high-level distributed data processing. Figure 3b shows the basic interrelationship between the extended data processing request, domain library and data processor (Java module). Domain library includes a hierarchy of data structures including atomic data structure, data structure describing single files, and high-level data structures which can be distributed over the nodes of storage. The DSL query can include statements with structures of different level. As a result there can be a set of high-level data structure which can be processed only during the aggregation process (they can be constructed using partial data structures prepared on each node of storage). So the prepared data analyzing procedures can be called both in map and reduce routines of distributed data processing.

# 4 Applications and Integration

This section shows an application example which is considered as a working example during the development of the technology and discusses the possible high-level e-Science solutions to be developed using the proposed technology.

## 4.1 Cyclones Path Analysis Application

The task of cyclones' paths analysis has significant importance for the protection Saint-Petersburg from storm surges (Averkiev & Klevanny, 2007). E.g. Gudrun cyclone (see Figure 4a and work on analysis of this cyclone (Suursaar, Kullas, Otsmann, Saaremäe, Kuik, & Merilain, 2006)) had appeared in January 2005 near the Ireland and caused a very dangerous flooding (239 cm) in Saint-Petersburg on January 9. To analyze the relevance between dangerous floods and paths of the cyclones a lot of data from various sources (including public archives of measurements and forecasts of pressure, wind, sea level etc.) can be analyzed and used for detailed simulation. Moreover the investigation on the new coming data, searching for similar historical data combined with simulation allows us to estimate possible dangerous level of coming cyclone.

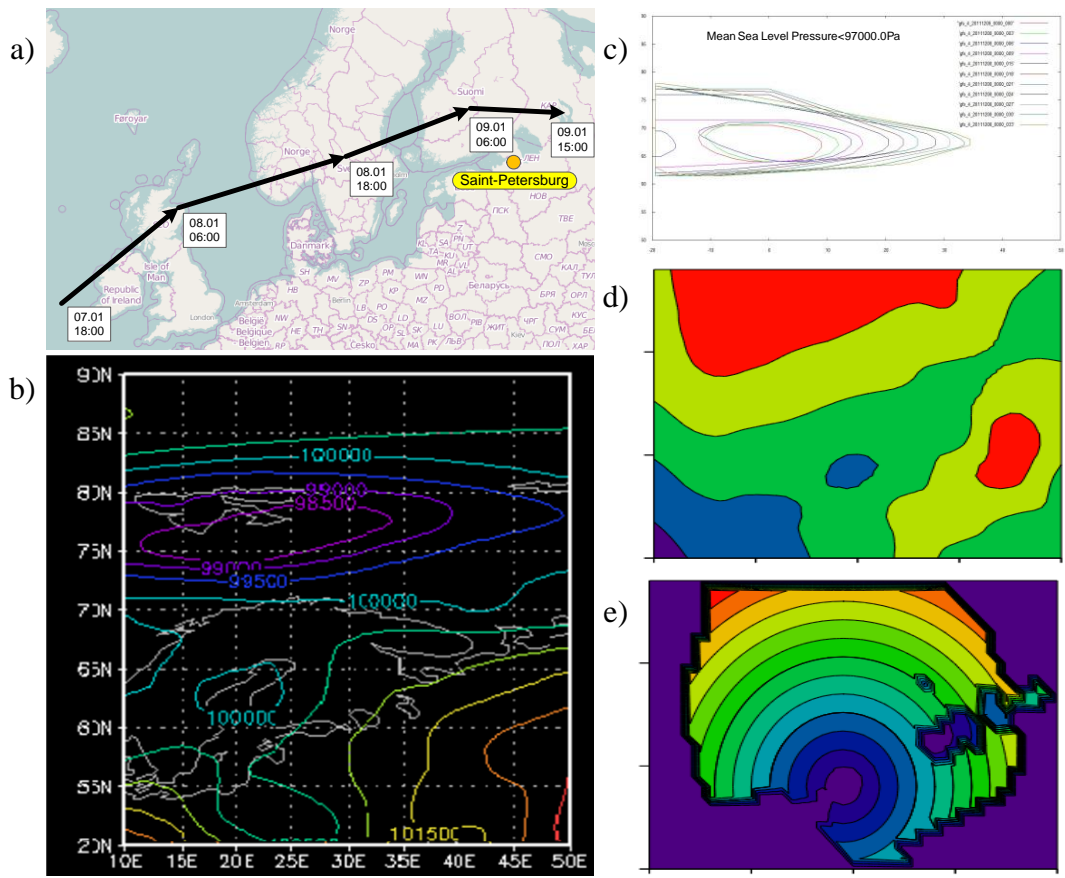

**Figure 4:** Cyclone paths analysis: a) Gudrun cyclone in January 2005; b) pressure field from GFS source; c) cyclone evolution; d) cyclone structure with sparse grid; e) cyclone structure with dense grid

To implement such application using the developed technologies several data sources are combined (e.g. GFS (Global Forecast System) – see Figure 4b). Analysis of historical data from such data sources can give the evolution patterns of the cyclones (see e.g. fig 4c for evolution of the cyclone over the time). After interpolation for the dense grid (see Figure 4d-4e) it is possible to obtain parametric description of the cyclone and its evolution over the time (including moving of the cyclone through the space).

The parameters of the cyclone can be used to generate an ensemble of possible cyclone evolutions. After that each of the cyclones within the ensemble can be used to simulate flooding impact within the defined conditions. As the initial dataset (even from single GFS source) can contain a lot of data the task (Terabytes), the problem can be considered as a BigData task. Moreover the data analysis task should be combined with the simulation task. Within the proposed approach the simulation can be performed on the remote nodes without moving data over the network.

A sketch (the actual DSL structure is now under development) example DSL query script is presented in the Figure 5. Here the keywords of the DSL are shown using bold font. Domain-specific keywords, values and constants are underlined. The request selects the cyclones with north-west direction and performs simulation of Baltic Sea water level at the end of cyclone path in the defined area. The result of this script should be the level trends in Saint-Petersburg for all found cyclones. The interpreter of DSL automatically identifies domain objects (cyclone paths), comparison procedure from domain library (compare by average direction), and simulation run of software package (Baltic Sea Model – BSM) with output parameters (sea level).

```
area 48.3416,-24.7851 - 66.1605,32.8710
time 01.01.2011 - 31.12.2011

select cyclon-path
  directon north-east
  out(Params[EndTime])

simulate
  with BSM
  semantic_association yes
  in(startTime: EndTime - 48h)
  out(level[440,414])
```

**Figure 5:** Proposed sketch of DSL query for cyclones paths analysis

## 4.2 BigData and Evolutionary Computing

This section discusses the possible further developments of the technology and its application within simulation-based e-Science solutions. One of the possible directions of complex data solution development is the combination of different model types within system-level analysis. E.g. BigData analytics solution can be used in combination with evolutionary computing to extend data-intensive scientific discovery (see Figure 6). In this case BigData analytics is used for basic classification and analysis of possible system evaluation scenarios. On the next stage the obtained set is extended with evolutionary computing. The evolution step can be combined with interactive simulation and assessment of the system state. Considering the population of possible scenario (a set of system configuration evolved within the evolutionary computing approach) the new states or/and development scenarios of the system can be discovered.

Figure 7 shows the implementation of such scenario for urban environment simulation-based investigation. Urban environment is considered as a complex system which includes multilayered networks of agents acting within the environment, layers of maps and relation with the global environment. A lot of the objects within urban environment can be described with large datasets: observations, crawling results, monitoring, forecasting etc. All these results can be used to identify classes of agents within urban environment, and building system-level simulation environment to discover the new agent classes, scenarios, structures in semi-automatic way. This approach can be used within Global System Science tasks (EUNOIA Consortium, 2013) for integrative exploration of complex environment including urban social systems simulation.

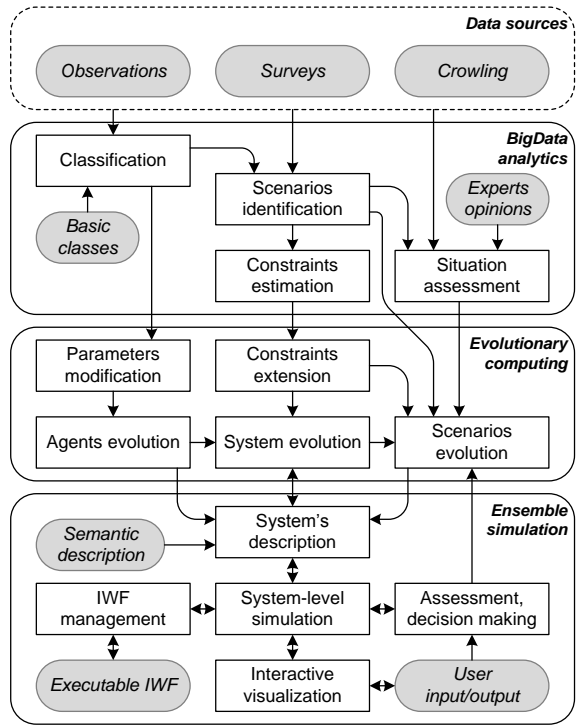

**Figure 6:** BigData analytics for evolutionary computing

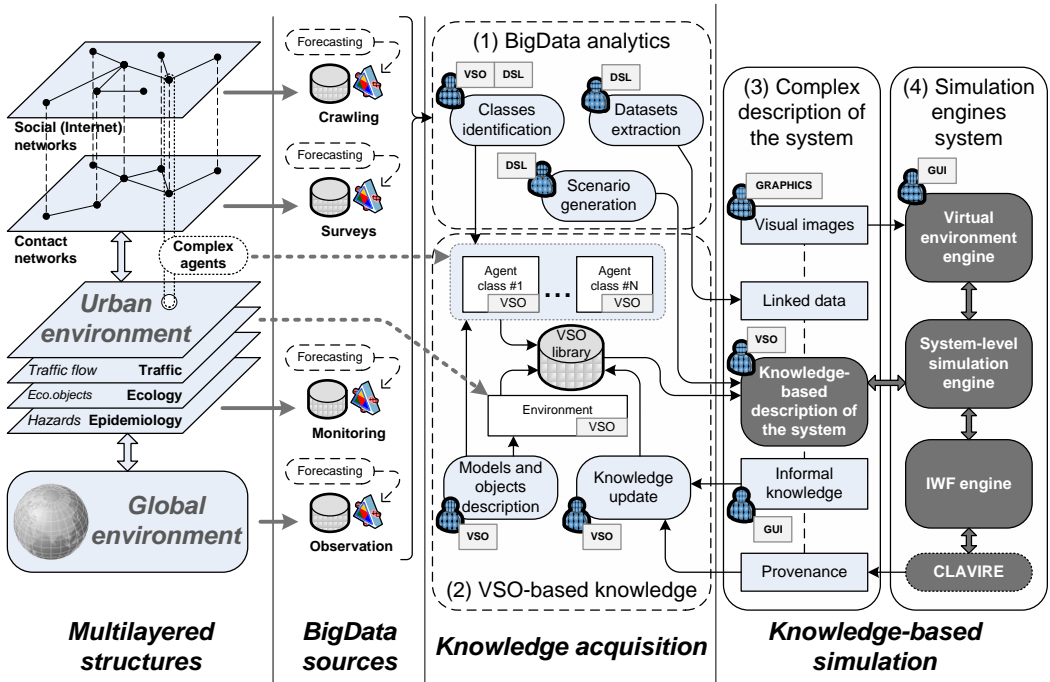

**Figure 7:** Urban environment investigation

# 5  Conclusion

The technology of dynamic DSL-based integration of BigData analytics tasks within cloud computing environment is aimed towards the implementation of a new high-level way of building scientific analytics and data-intensive discovery tasks, where all the infrastructure specifics is hidden behind the automatic performed procedures. It can serve as a platform for Model-as-a-Service approach (Krishna & Varma, 2012) where the core provided resource is a model available for composition within analytics tasks. It can be used within Global System Science area where integration of different data resources and high-level analytics tasks plays an important role. We believe that proposed technology, within further development can extend the semantic-based solutions and offer the domain user a flexible tool for data-intensive scientific discovery.

The presented DSL-based technology for high-level analytic tasks' definition is now being developed and future plans include full implementation of proposed approaches and techniques as well as integration of the technology into more complicated solutions.

*Acknowledgements:* This work was financially supported by Government of Russian Federation, Grant 074-U01; Grant for the Netherlands Institute for Advanced Study in the Humanities and Social Sciences (NIAS); and project "Big data management for computationally intensive applications" (project #14613).

# References

*Apache Hadoop*. (n.d.). Retrieved from Apache: http://hadoop.apache.org/

*Apache OODT*. (n.d.). Retrieved from Apache: http://oodt.apache.org/

Assuncao, M. D., Calheiros, R. N., Bianchi, S., Netto, M. A., & Buyya, R. (2013). Big Data Computing and Clouds: Challenges, Solutions, and Future Directions. *arXiv preprint* (arXiv:1312.4722).

Averkiev, A. S., & Klevanny, K. A. (2007). Determining cyclone trajectories and velocities leading to extreme sea level rises in the Gulf of Finland. *Russian Meteorology and Hydrology , 32* (8), 514-519.

Baranowski, M., Belloum, A., & Bubak, M. (2013). MapReduce Operations with WS-VLAM Workflow Management System. *Procedia Computer Science , 18*, 2599–2602.

Dearle, F. (2010). *Groovy for Domain-Specific Languages.*

Doulkeridis, C., & Nørvåg, K. (2013, June). A survey of large-scale analytical query processing in MapReduce. *The VLDB Journal* , 1-26.

EUNOIA Consortium. (2013). *Global Systems Science and Urban Development.*

Fiore, S., D'Anca, A., Palazzo, C., Foster, I., Williams, D. N., & Aloisio, G. (2013). Ophidia: Toward Big Data Analytics for eScience. *Procedia Computer Science , 18*, 2376-2385.

Foster, I., & Kesselman, C. (2006). Scaling system-level science: Scientific exploration and IT implications. *IEEE Computer , 39* (11), 31-39.

*Global Forecast System*. (n.d.). Retrieved from National Centers for Environmental Prediction: http://www.emc.ncep.noaa.gov/index.php?branch=GFS

Hey, T., Tansley, S., & Tolle, K. (2009). *The fourth paradigm: data-intensive scientific discovery.*

Knyazkov, K. V., Kovalchuk, S. V., Tchurov, T. N., Maryin, S. V., & Boukhanovsky, A. V. (2012). CLAVIRE: e-Science infrastructure for data-driven computing. *Journal of Computational Science , 3* (6), 504-510.

Kovalchuk, S. V., Razumovskiy, A. V., & Spivak, A. I. (2013). Domain-specific technology for big data storing and processing within cloud computing platform CLAVIRE. *Dinamika Slozhnyh Sistem (in Russian)* (3), 106-109.

Kovalchuk, S. V., Smirnov, P. A., Knyazkov, K. V., Zagarskikh, A. S., & Boukhanovsky, A. V. (2013). Knowledge-based Expressive Technologies within Cloud Computing Environments. *Proceedings of the 8th International Conference on Intelligent Systems and Knowledge Engineering*, *(in press), arXiv preprint (arXiv:1312.7688)*.

Kovalchuk, S. V., Smirnov, P. A., Kosukhin, S. S., & Boukhanovsky, A. V. (2012). Virtual Simulation Objects Concept as a Framework for System-Level Simulation. *IEEE 8th International Conference on E-Science*, (pp. 1-8).

Krishna, P. R., & Varma, K. I. (2012). *Cloud analytics: A path towards next generation affordable BI.* White paper.

McPhillips, T., Bowers, S., Zinn, D., & Ludäscher, B. (2009). Scientific workflow design for mere mortals. *Future Generation Computer Systems , 25* (5), 541-551.

Philip Chen, C. L., & Zhang, C.-Y. (2014). Data-Intensive Applications, Challenges, Techniques and Technologies: A Survey on Big Data. *Information Sciences , (in press)*.

Suursaar, Ü., Kullas, T., Otsmann, M., Saaremäe, I., Kuik, J., & Merilain, M. (2006). Cyclone Gudrun in January 2005 and modelling its hydrodynamic consequences in the Estonian coastal waters. *Boreal Environment Research , 11* (2), 143-159.

van Deursen, A., Klint, P., & Visser, J. (2000). Domain-Specific Languages: An Annotated Bibliography. *ACM SIGPLAN Notices , 35* (6), 26-36.